\documentclass[]{article}
\usepackage{plain}
\usepackage{fullpage}
\usepackage{subfigure}
\usepackage[normalem]{ulem}
\usepackage{siunitx}
\usepackage{wrapfig, booktabs}
\usepackage{multicol}
\usepackage[colorinlistoftodos]{todonotes}
\usepackage{multirow}
\usepackage{hhline}
\usepackage{url}
\graphicspath{{./figures/}}
\usepackage{graphicx}

\usepackage{amssymb}
\usepackage{amsmath}
\usepackage{float}
\usepackage{tabularx}
\usepackage{caption}
\usepackage{rotating}
\usepackage{dirtree}

\newcommand{\eat}[1]{}
\usepackage[english, status=draft,inline=false,margin=true]{fixme}
\fxusetheme{color}
\usepackage{tikz}
\usetikzlibrary{shapes,arrows,snakes,backgrounds} 
\usepackage{smartdiagram}

\title{A Comparison of Statistical and Machine Learning Algorithms for Predicting Rents in the San Francisco Bay Area}
\author{%
  \textbf{Paul Waddell}\\
  waddell@berkeley.edu \\
  \hfill\break%
  \textbf{Arezoo Besharati-Zadeh}\\
  arezoo.bz@berkeley.edu \\
  \hfill\break%
}

\begin{document}
\maketitle

\begin{abstract}
Urban transportation and land use models have used theory and statistical modeling methods to develop model systems that are useful in planning applications.  Machine learning methods have been considered too 'black box', lacking interpretability, and their use has been limited within the land use and transportation modeling literature.  We present a use case in which predictive accuracy is of primary importance, and compare the use of random forest regression to multiple regression using ordinary least squares, to predict rents per square foot in the San Francisco Bay Area using a large volume of rental listings scraped from the Craigslist website.  We find that we are able to obtain useful predictions from both models using almost exclusively local accessibility variables, though the predictive accuracy of the random forest model is substantially higher.

\end{abstract}

\newpage

\section{Introduction}
\label{sec:introduction}

Development of urban transportation and land use models has traditionally relied extensively on domain knowledge, theory, and statistical methods such as multiple regression and discrete choice models.  Although machine learning methods have been available for many years and demonstrated to produce more accurate predictions than statistical models such as multiple regression, they have not been widely adopted within the urban modeling literature.  One of the main reasons for this is that in research using statistical models (whether frequentist or Bayesian), the applications are often motivated by a need to be able to interpret the coefficients of the model within the context of domain theory about their sign and significance. By contrast, researchers used to statistical modeling paradigms have been wary about the perceived lack of interpretability of models developed using machine learning methods such as neural networks.  Further, models developed for planning or policy applications are often motivated by a need to undertake counter-factual analysis of the potential impacts of different policy inputs, in order to undertake ex-ante evaluation of the policies.  This requires some degree of causal inference, or at at least a model with a theoretical structure that the researcher can argue is suitable for counter-factual analysis.  By contrast, again, machine learning methods tend to focus on maximizing the predictive accuracy rather than on counter-factual analysis for policy or planning.  

In this paper we examine a use case that lends itself to the use of machine learning methods, since the predictions are used mainly to bootstrap a structural model.  The application is hedonic modeling of rents, to be used as starting values for a model that is a structural microsimulation of demand and supply of housing, and which incorporates a short-term market clearing component that adjusts prices until the demand for housing would clear all submarkets -- meaning that predicted demand is less than or equal to available supply in all submarkets. For this purpose, it is valuable to obtain the most accurate possible initial prediction of rents or prices, since that predicted value will influence the demand predictions, and a poor prediction of prices or rents will generate a lower quality prediction of demand.  If the estimated parameters of the demand model were sufficiently robust with respect to price and amenities of housing, then one might hope that the market clearing algorithm would adjust prices to more accurately reflect true demand.  But in the presence of poor predictions of prices, one might have less confidence that the estimated parameters of the demand model are sufficiently robust.  More accurate price and rent predictions should help achieve robust estimation results from the demand model, and more efficient convergence of the market clearing algorithm.

We develop a hedonic regression model of rent per square foot, first using Ordinary Least Squares regression \cite{greene-2002}, and subsequently using random forest regression, a decision tree method within machine learning \cite{breiman-1984,breiman2001}.  The literature of hedonic modeling of housing prices is voluminous, dating to at least the work of Griliches on the automobile market \cite{griliches1961hedonic}, and early application to modeling housing rents \cite{gillingham1970hedonic}. The theoretical formulation of hedonic modeling is generally attributed to Rosen \cite{rosen-1974}, and is grounded in Lancaster's theory of consumer demand \cite{Lancaster-1966}. Housing prices and rents have also been examined previously using random forest regression, and compared to multiple regression, for example in the context of Ljubljana, Slovenia \cite{Ceh-2018}, and broader comparisons of multiple regression and random forest regression for evaluating variable importance are also available \cite{Gromping-2009}.  Our paper contributes to the small emerging literature on the use of machine learning methods such as random forest for analyzing housing prices and rents in the context of land use and transportation modeling.  It is also novel in using volunteered geographic information from Craigslist rental listings, leveraging prior work to scrape rental listings \cite{boeing-waddell-2017}.

\section{Case Study and Data}
\label{sec:CaseStudy}

The context of this study is the San Francisco Bay Area, with a population of over seven million and encompassing over one hundred municipalities across nine counties.  It is home to Silicon Valley and owing in part to its robust technology sector, it is the most expensive metropolitan housing market in the United States.    

\subsection{Data Sources}

For this study we scraped Craigslist rental listings from November 2016 through July 2018, and filtered and cleaned the data, adapting the methods used in \cite{boeing-waddell-2017}.  The result of the data collection and cleaning yielded over 350,000 rental listings which contained information on the listing date, the location (Latitude, Longitude), asking rent, square footage, number of bedrooms and number of bathrooms.  Since the objective of this project was to generate a rental model that could be used in an integrated microsimulation of the Bay Area real estate market at a building level, we used only the location, rent, and square footage information from the listings.

To augment the listing attributes we developed a series of locational attributes and associated them with the listings data.  We employed street networks representing the walking network and a driving network containing tertiary streets and higher capacity roads to measure accessibility, using the OSMNX library \cite{boeing-osmnx-2017} to create and clean the networks, and the Pandana library \cite{foti2012generalized} to compute localized accessibility measures.  We developed a synthetic population using Synthpop, a library adapted from PopGen \cite{ye-trb-2009}.  Data for parcels, buildings, and employment by address were obtained for the 9-County Bay Area from the Metropolitan Transportation Commission. We computed a series of localized accessibility measures on the walk and drive networks to provide localized and more regional context measures.   Each listing was assigned to its nearest node on the two networks, and each parcel and building was similarly assigned to its nearest node on each of the two networks.  The localized measures were generally computed within 500 meters as network distances, either as a simple sum, or an average of the variables of interest.

\subsection{Variables}
From the buildings database we computed the average residential square feet per residential unit for those buildings containing residential units.  We used the following attributes of households from the synthetic population: household income, household size, age of householder, presence of children, race of householder, and an indicator of whether the householder was Hispanic.  We also used data on jobs by location to compute accessibilities to employment within 500 meters, 1500 meters, 10 kilometers, and 25 kilometers, all measured as distances on the network, with those 3 kilometers or below measured on a walking network, and those above 3 kilometers measured on a driving network with tertiary streets or higher.

\begin{table}
{\scriptsize
\begin{center}
\caption{Statistical Profile of Variables After Clipping Outlier Values} 
\label{tab:data_after} 
\begin{tabular}{l r  r  r  r  r  r  r  r}

variable & count & mean & std & min & 25\% & 50\% & 75\% & max \\
\hline
rent\_sqft & 363010.0 & 3.0 & 1.0 & 0.0 & 2.0 & 3.0 & 4.0 & 11.0 \\
res\_sqft\_per\_unit & 363010.0 & 994.0 & 430.0 & 212.0 & 710.0 & 904.0 & 1150.0 & 3600.0 \\
units\_500\_walk & 363010.0 & 664.0 & 662.0 & 0.0 & 193.0 & 437.0 & 876.0 & 2317.0 \\
sqft\_unit\_500\_walk & 363010.0 & 1455.0 & 712.0 & 0.0 & 1059.0 & 1436.0 & 1803.0 & 3699.0 \\
rich\_500\_walk & 363010.0 & 133.0 & 148.0 & 0.0 & 27.0 & 81.0 & 166.0 & 528.0 \\
singles\_500\_walk & 363010.0 & 201.0 & 254.0 & 0.0 & 35.0 & 101.0 & 228.0 & 868.0 \\
elderly\_hh\_500\_walk & 363010.0 & 92.0 & 102.0 & 0.0 & 21.0 & 56.0 & 117.0 & 363.0 \\
children\_500\_walk & 363010.0 & 226.0 & 189.0 & 0.0 & 79.0 & 186.0 & 327.0 & 755.0 \\
jobs\_500\_walk & 363010.0 & 759.0 & 1295.0 & 0.0 & 43.0 & 220.0 & 748.0 & 5247.0 \\
jobs\_1500\_walk & 363010.0 & 6589.0 & 8770.0 & 0.0 & 1206.0 & 3110.0 & 7220.0 & 32501.0 \\
jobs\_10000 & 363010.0 & 165285.0 & 117970.0 & 0.0 & 74380.0 & 127551.0 & 236962.0 & 412326.0 \\
jobs\_25000 & 363010.0 & 498022.0 & 229898.0 & 37.0 & 322181.0 & 584284.0 & 696465.0 & 787748.0 \\
pop\_10000 & 363010.0 & 333207.0 & 191209.0 & 0.0 & 183445.0 & 300216.0 & 459446.0 & 763247.0 \\
pop\_black\_10000 & 363010.0 & 14010.0 & 18451.0 & 0.0 & 2709.0 & 5754.0 & 20794.0 & 90219.0 \\
pop\_hisp\_10000 & 363010.0 & 57468.0 & 42489.0 & 0.0 & 27776.0 & 45772.0 & 81072.0 & 201053.0 \\
pop\_asian\_10000 & 363010.0 & 106511.0 & 77819.0 & 0.0 & 37199.0 & 93097.0 & 175019.0 & 282688.0 \\
\hline 
\end{tabular}
\end{center}
}
\end{table}


\section{Methods}
\label{sec:methods}

\subsection{Data Pre-Processing}
Prior to estimating the models on training subsamples of our data, we examined the data to identify potential problems with the data that might adversely impact the quality of the models.  In particular, outliers are well known to influence model parameters in ordinary least squares regression, so we used the Pandas clip function to recode values above the 99th percentile on all accessibility variables in order to reduce their impact on the model.

We also examined the distribution of the variables, and used log transformations to normalize them.  Most had significant skewness prior to the transformation.  We did not, for purposes of this paper, undertake further data cleaning, in order to simplify the exposition and focus on the main objective of the paper to compare two very different methods to predict rental prices.

\subsection{Hedonic Regression with Ordinary Least Squares}

The hedonic price model we estimate using Ordinary Least Squares (OLS) method of minimizing the sum of the squared errors is represented as $Y = f(S\beta, N\gamma)+ \epsilon$,  where $Y$ is a vector of rents per square foot in our rental listings, $S$ is a vector of structural characteristics, and $N$ is a vector of neighborhood and accessibility characteristics surrounding each rental listing.  This can be specified as a linear model $Y = X\beta + \epsilon$, with $\epsilon$ assumed to be independently and identically distributed (\emph{iid}).  Using this assumption on the distribution of $\epsilon$, we can estimate the $\beta$ parameters using OLS by computing $\hat{\beta} = (X^TX)^{-1}X^TY$.
We use the StatsModels Python library \cite{seabold2010statsmodels} compute the model estimation and predictions for model evaluation.

\subsection{Hedonic Regression with Random Forest}

The development of Random forests as an ensemble classification and regression approach by Breiman \cite{breiman2001} has produced considerable interest owing to its robust predictive capabilities and minimal tuning requirements.  We summarize the method here using the exposition of \cite{segal-2003}. A random forest is a collection of tree predictors $h(x;\theta_k), k=1,...,K$ where $x$ i an input vector of length $p$, and random vectors $X$ and $\theta_k$ are independently and identically distributed (\emph{iid}). We subset the data into a training and a testing subsample, and take independent draws from the training data, which is a joint distribution of $(\boldsymbol{X}, Y)$. The random forest regression prediction is an average over the collection $\bar{h}(x) = (1/K)\sum_{k=1}^Kh(x;\theta_k)$.  We use the Scikit-Learn library \cite{pedregosa-2011} to train the random forest model on our data.

\section{Model Estimation on Training Sample}
\label{sec:results}

In this section we present results from estimating the multiple regression model using ordinary least squares, and training the random forest regression. One of the standard practices used in machine learning is to split the observed data into training and testing samples, and to use only the training sample to train the model.  Both for consistency, and to adopt the valuable practice of out-of-sample validation, we use the same approach for the OLS regression as we do for training the random forest regression: we split the observed data, using two thirds of the data for estimation (training) and separating one third of the data to use for out-of-sample prediction and cross-validation.

\subsection{Ordinary Least Squares}

We used the training sample of 242,000 rental listings to estimate a hedonic model on rent per square foot from the Craigslist rental listings data collected from November 2016 through July, 2018 for the San Francisco Bay Area.  As is common in the literature, we used a log transformation of the dependent variable, and also log-transformed all explanatory variables. 

\begin{table}[H]
\label{tab:table2}
\caption{Results of OLS Estimation}
\end{table}

\begin{figure}[H]
\centering
\includegraphics[width=16cm]{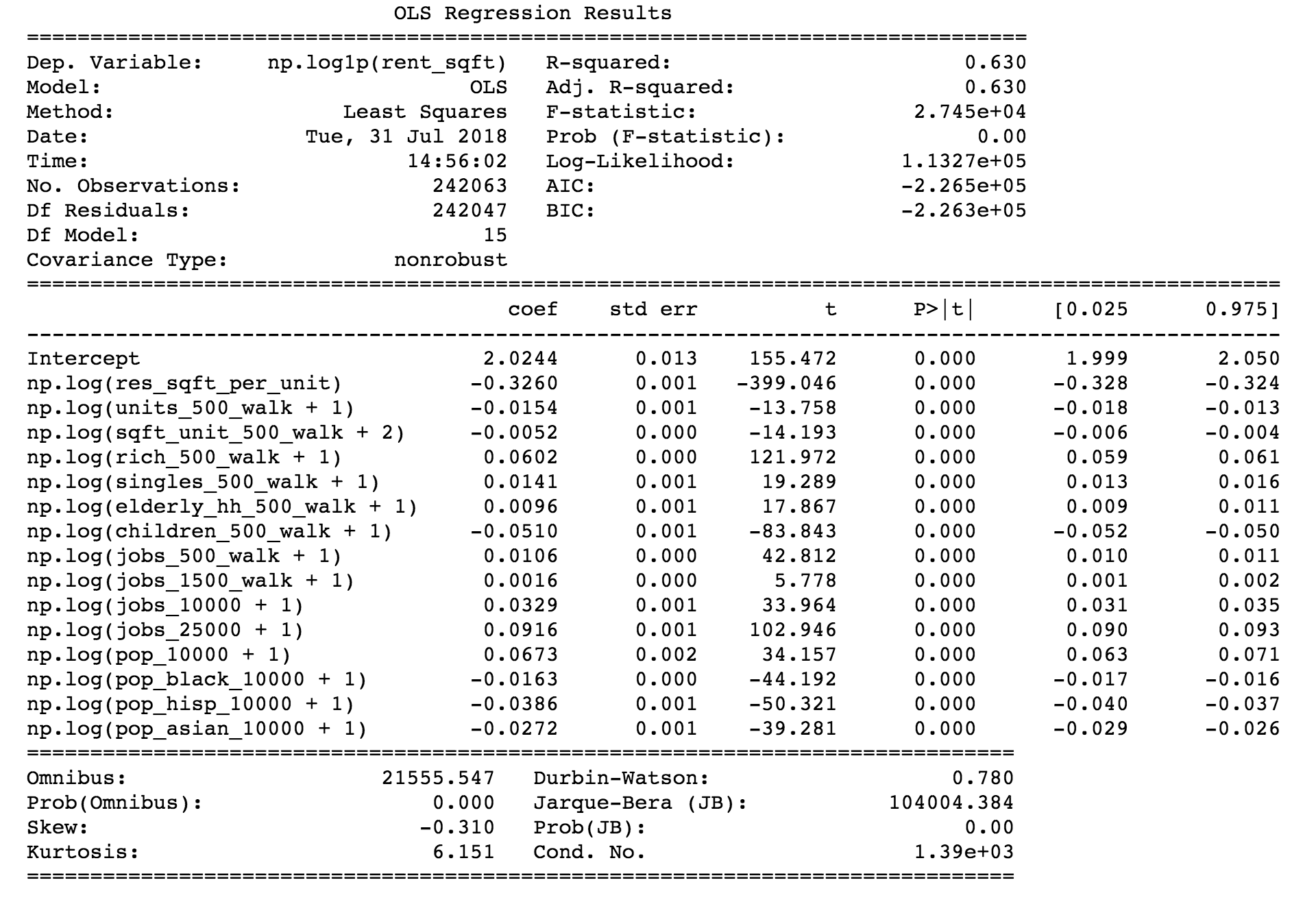}
\vspace{10pt}
\end{figure}

The estimated model appears to fit the training data reasonable well considering that the only information included about the unit besides accessibility variables is is square footage.  Even with such limited numbers of attributes, the model has an Adjusted R-squared of 0.63.  The Mean squared error is= 0.45, and the RMSE is 0.67.  Key variables have the correct sign and are significant.  For clarity of exposition we will not repeat that all variables are log transformed, which also lends itself to a straightforward interpretation of the coefficients as elasticities, with a one percent change in an explanatory variable being associated with a percentage change in the rent per square foot as indicated by its estimated coefficient.  

Keeping in mind that the dependent variable is expressed as a monthly rent per square foot, the size of the units in square feet is negatively associated with rent per square foot, consistent with a diminishing marginal utility as square footage increases.  The density of housing within 1/2 kilometer is positively associated with rent per square foot, and the size of units in square feet within 1/2 kilometer is negatively correlated with rent per square foot.  The number of households within 1/2 kilometer with incomes above \$150,000, the number of single-person households, and the number households with a householder of age over 65 are all positively correlated with rent per square foot.  Number of households with children within 1/2 kilometers on the other hand, is negatively correlated with rent per square foot.  Jobs within 1/2 kilometer, 1.5 kilometer, 10 kilometers and 25 kilometers are all associated with higher rents per square foot, with growing significance and elasticity as the distance threshold increases.  Finally, the number of householders within 10 kilometers who identify their race as Black, Hispanic, or Asian, are negatively correlated with rent per square foot.

\subsection{Random Forest}

\begin{figure}[H]
\centering
\includegraphics[width=16cm]{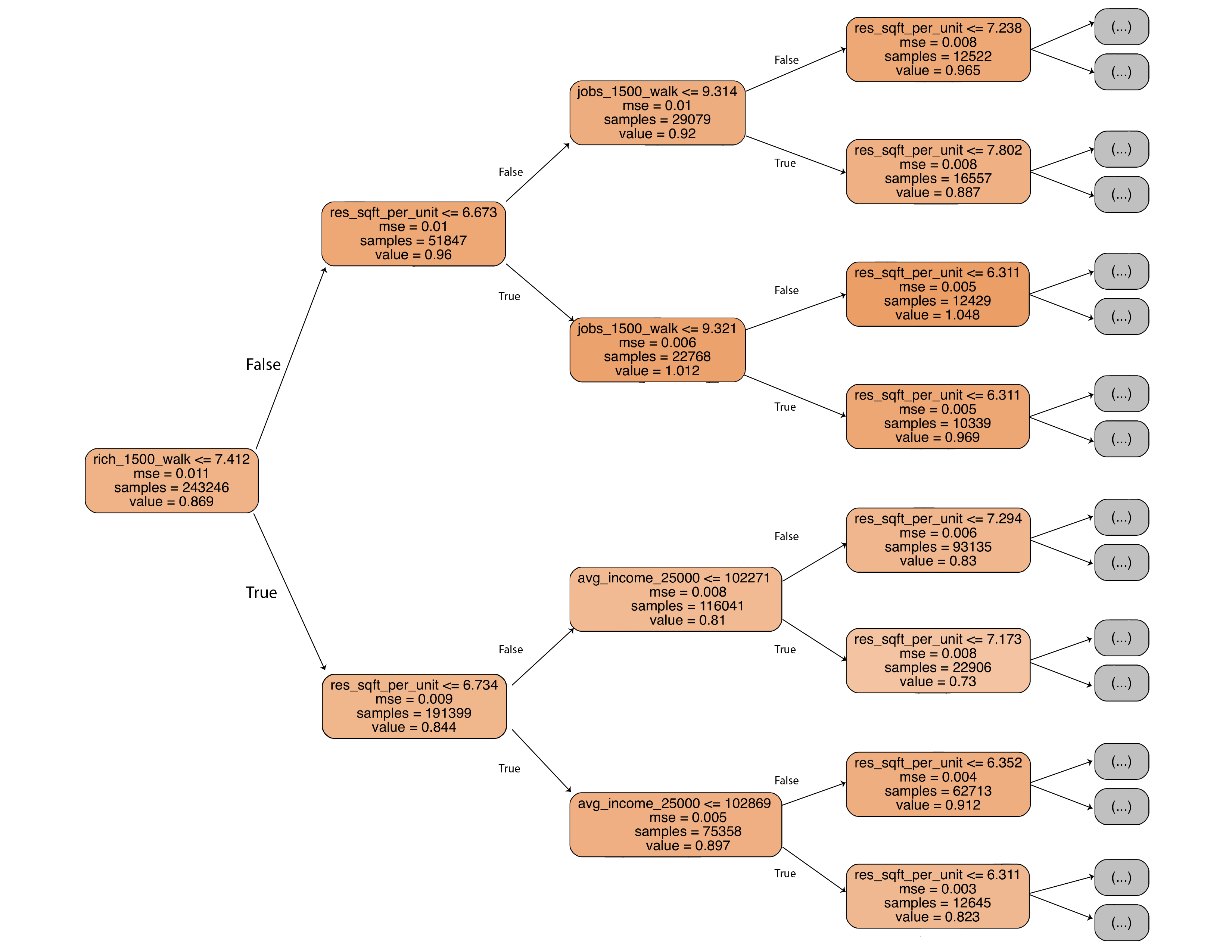}
\caption{\label{fig:tree}One tree from random forest ensemble of regression trees}
\end{figure}

We turn next to the random forest model and examine its training results on the same training dataset, using the same variables used in the OLS regression.  Results from random forest training are quite different than those from OLS estimation, since the underlying algorithms are fundamentally different.  Random forest leverages regression trees, and averages over an ensemble of regression trees to make its predictions.  The depth of the trees and the number of samples drawn to generate different trees are controlled by the researcher, and provide a means to tune the degree to which the model trades off bias and variance, in order to avoid over-fitting.

Figure \ref{fig:tree} represents one regression tree of the ensemble of regression trees generated by random forest on the training dataset. It is included to help illustrate how regression trees use decision trees to split variables recursively in order to capture nonlinear patterns within the data.

\begin{figure}[H]
\centering
\includegraphics[width=15cm]{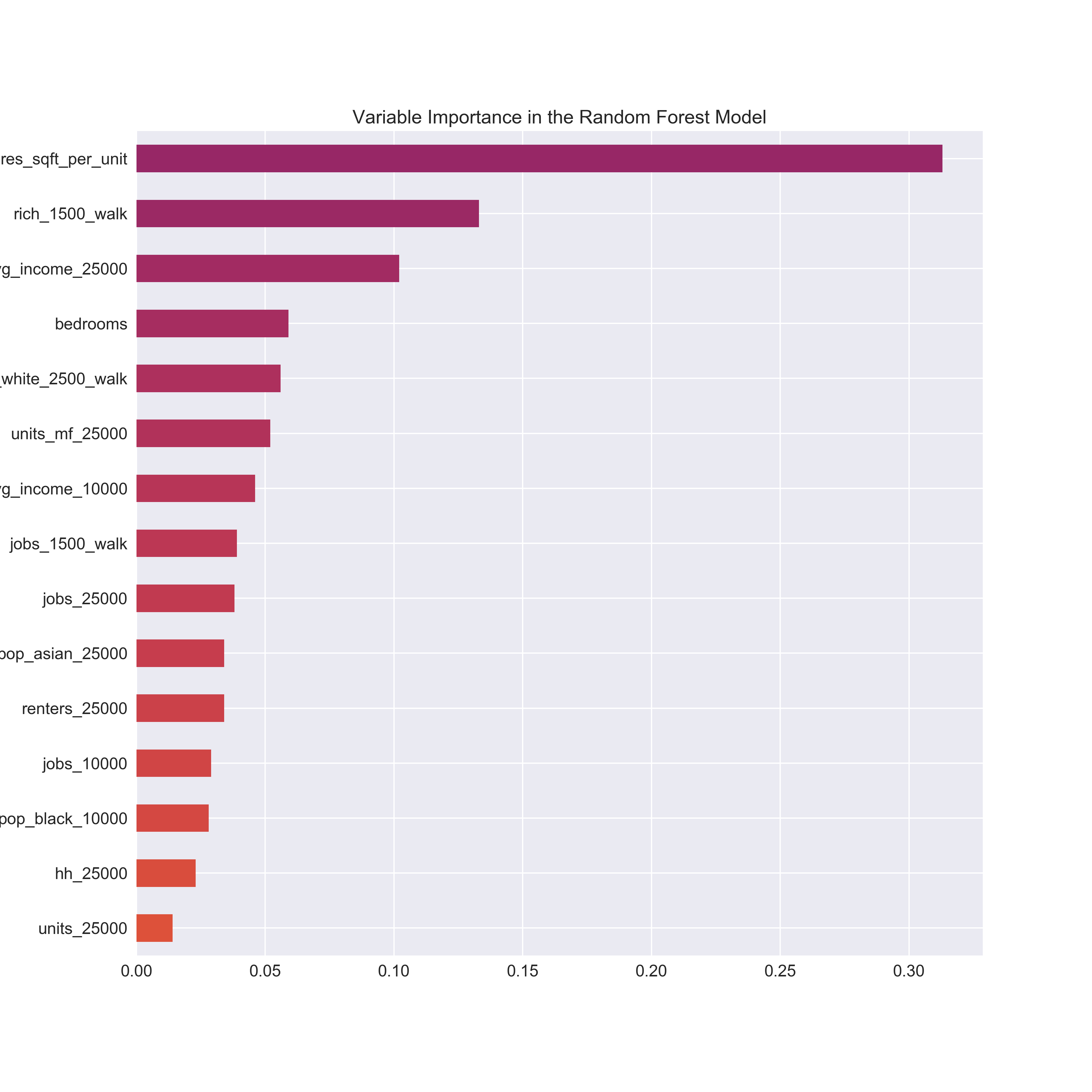}
\caption{\label{fig:rf_ranking}Variable Importance Ranking from Random Forest Regression}
\end{figure}

Figure \ref{fig:rf_ranking} depicts the relative importance of the most important variables contributing to its predictions.  It is not directly comparable to OLS coefficients, but it does provide insight into the which variables most influence the prediction, and in this way is more interpretable than some machine learning methods.  In terms of fit to the training data, the random forest model has an accuracy score on the training data of .96, which comparable to the R-squared of OLS.  The Mean squared error on the training data is= 0.0, and the RMSE is 0.02.

\section{Cross Validation Results}
\label{sec:cross-validation}

In this section we examine how both models performed when the estimated (trained) model is applied to a set of observations not used in training the models.  This practice is commonly used to evaluate whether over-fitting has occurred, a situation in which the model is excessively tuned to the training dataset and does not perform very well when it is generalized to other data.  For this purpose, we use the 1/3 of the original data that was split into a testing dataset.  It shares no observations with the training data.

\begin{figure}[H]
\centering
\subfigure[Ordinary Least Squares]{\includegraphics[width=8cm]{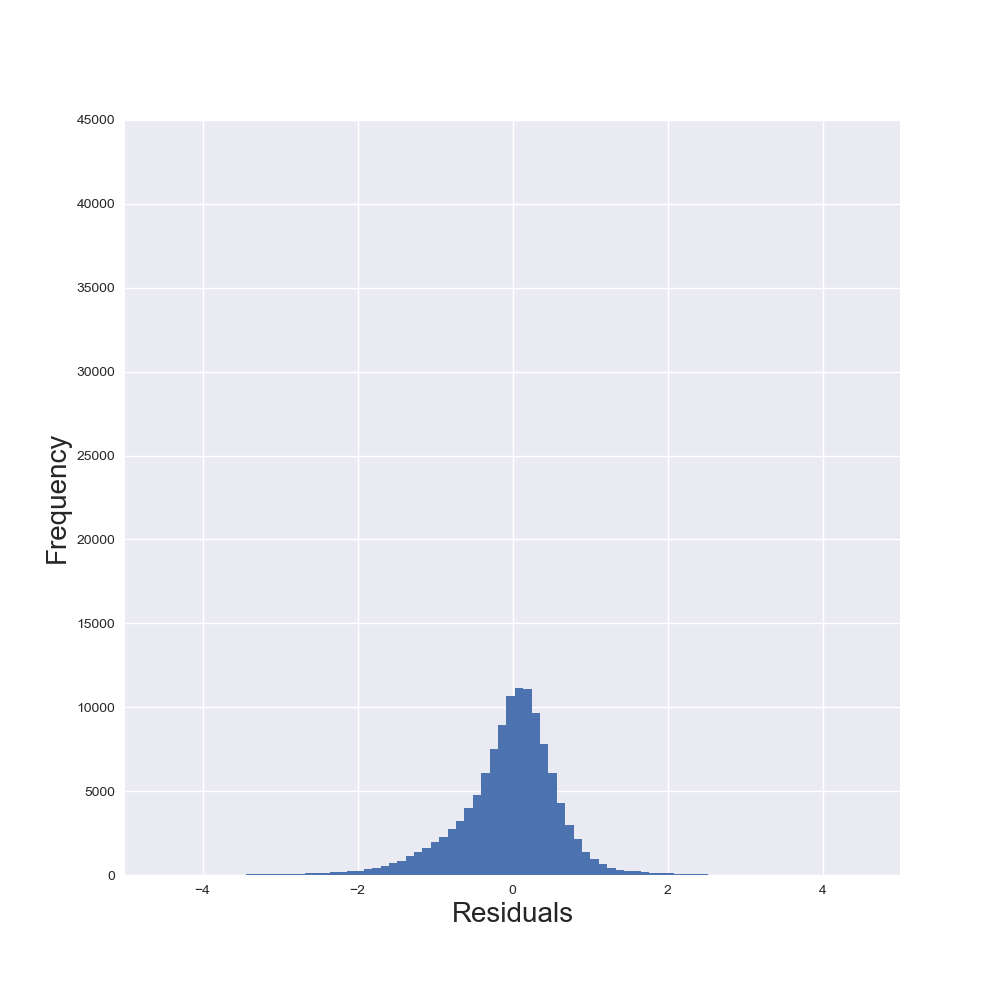}}
\subfigure[Random Forest]{\includegraphics[width=8cm]{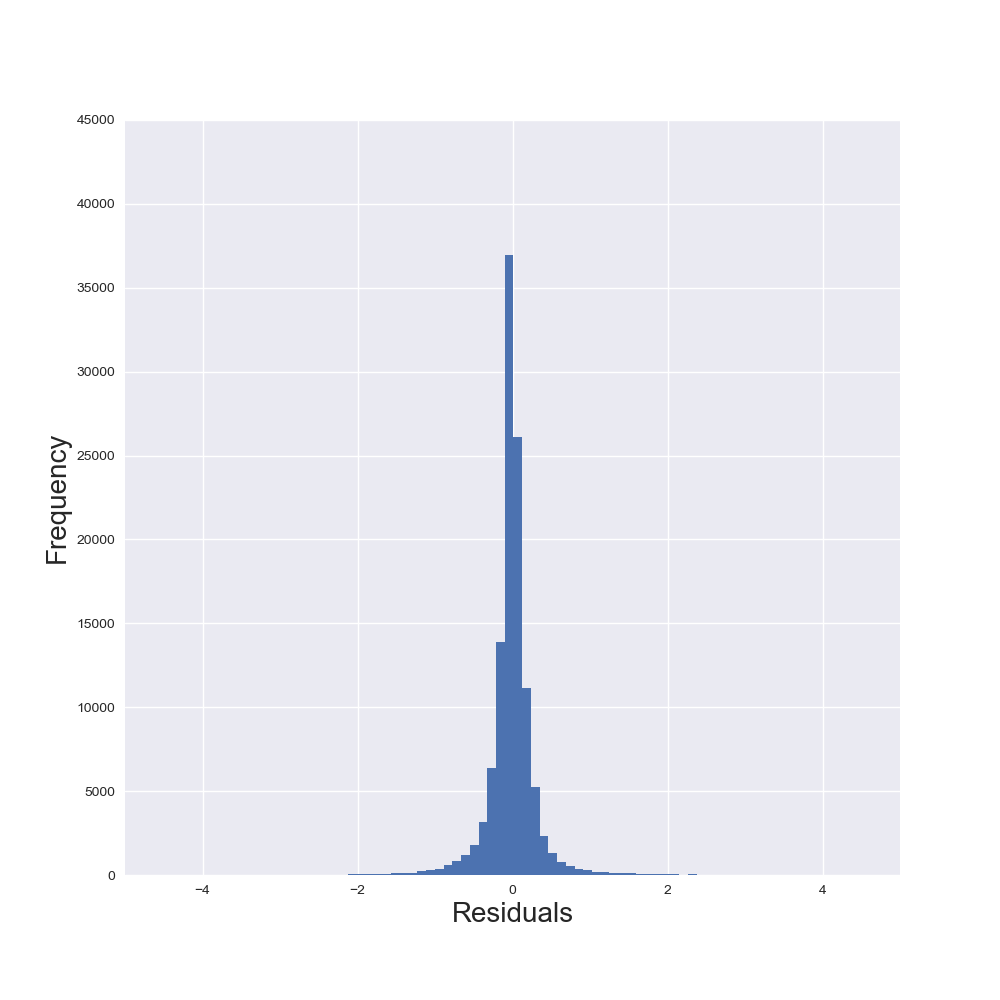}}
\caption{Distribution of Residuals of OLS and Random Forest}
\label{fig:normal}
\end{figure}

The first result we examine is a comparison of the distribution of the residuals when we predict the model on these new data and compare the predicted values to the observed values of rent per square foot.  These results are shown in Figure \ref{fig:normal}.  While both plots indicate little to no bias in the models, the random forest plot clearly shows a much lower variance of the residuals, indicating a superior set of out of sample predictions compared to the predictions from OLS.

\begin{figure}[H]
\centering
\subfigure[Ordinary Least Squares]{\includegraphics[width=7.5cm]{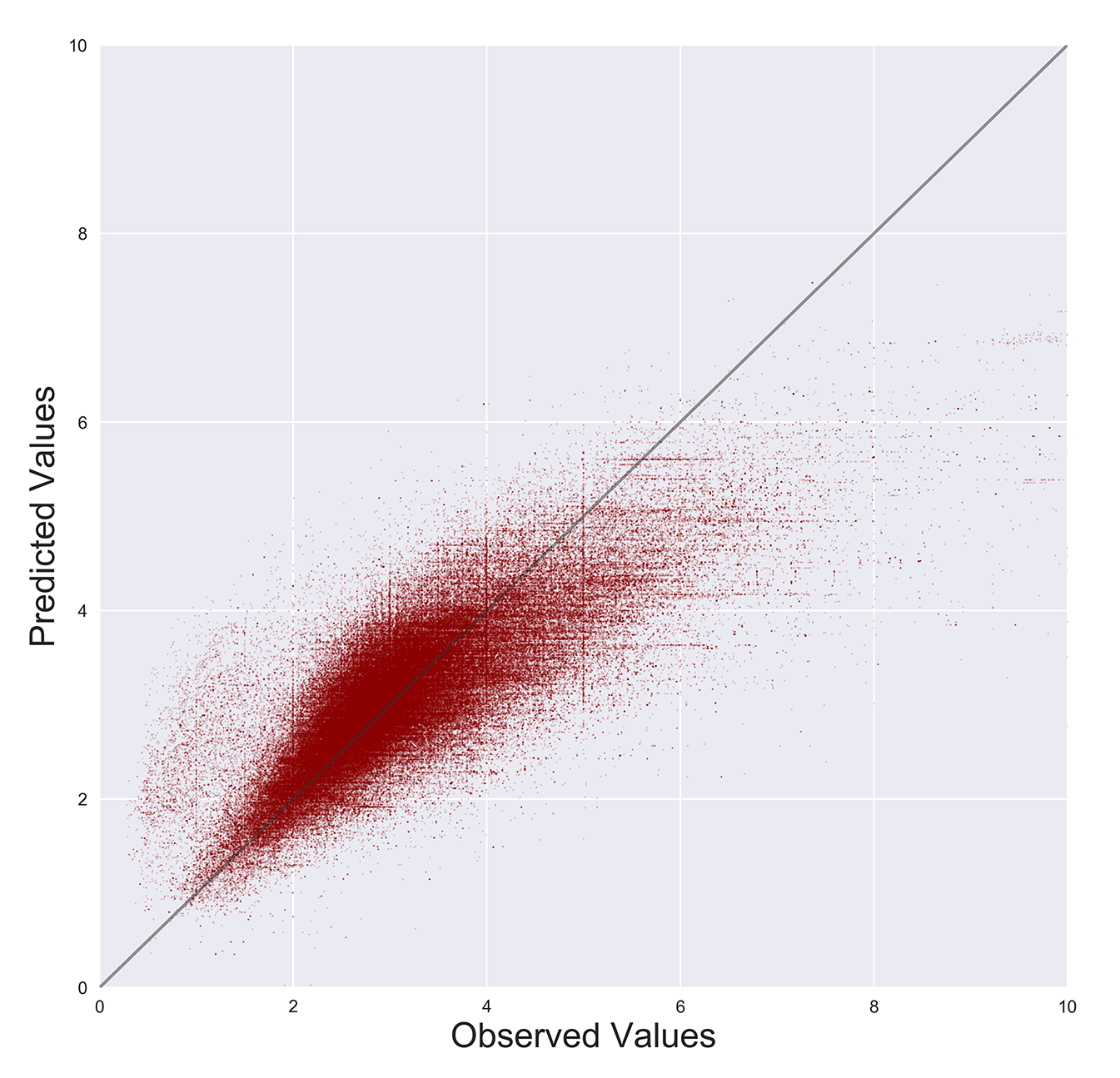}}
\subfigure[Random Forest]{\includegraphics[width=7.5cm]{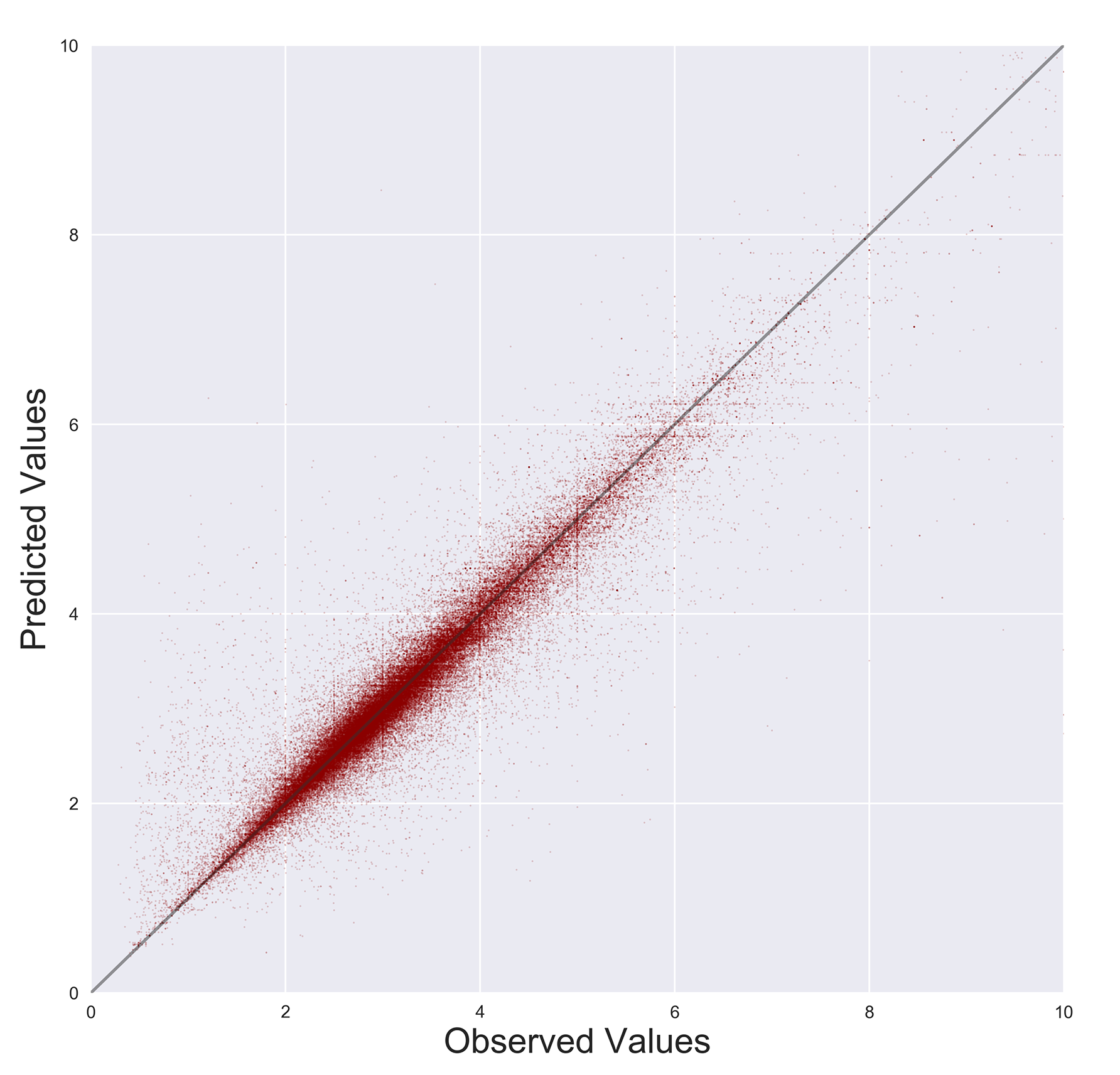}}
\caption{Predicted vs Observed Values for OLS and Random Forest}
\label{fig:predobs}
\end{figure}

Figure \ref{fig:predobs} displays predicted values plotted against observed values for the test dataset for both models.  Again it is clear that the random forest model fits the test data much better than the predictions of the model estimated with OLS.  The predictions of random forest  map closely to the 45 degree line of a perfect fit, and show no signs of distorion at lower or higher values of the observed data.  By contrast, the OLS predictions show some artifacts, with a cloud of points being over-predicted at low ranges of observed rent per square foot, while at high ranges of observed rent per square foot the OLS predictions appear to be skewed downwards, towards under-prediction.  The higher dispersion of the errors in the OLS predictions are also  evident.

\begin{figure}[H]
\centering
\subfigure[Ordinary Least Squares]{\includegraphics[width=7.5cm]{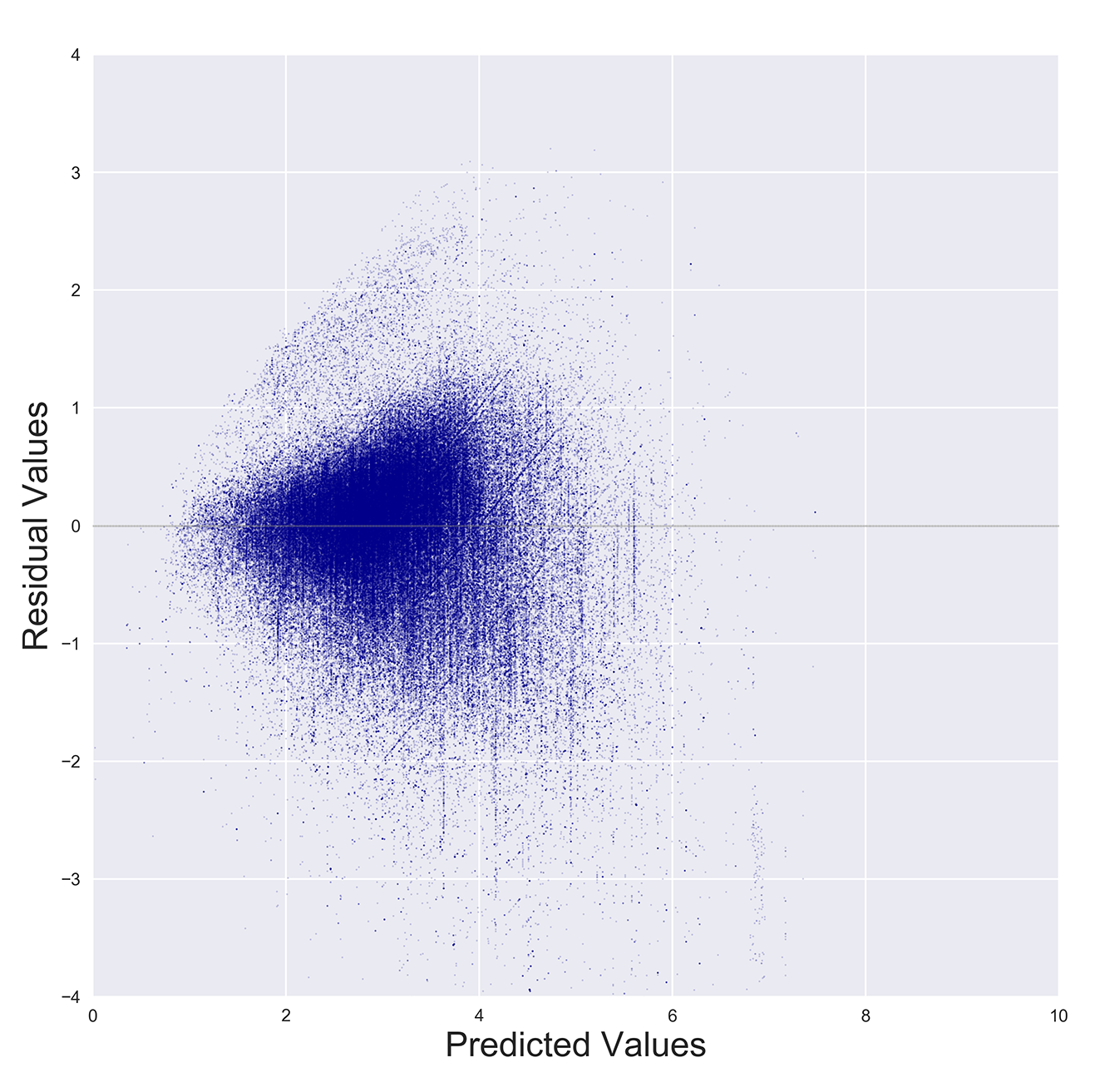}}
\subfigure[Random Forest]{\includegraphics[width=7.5cm]{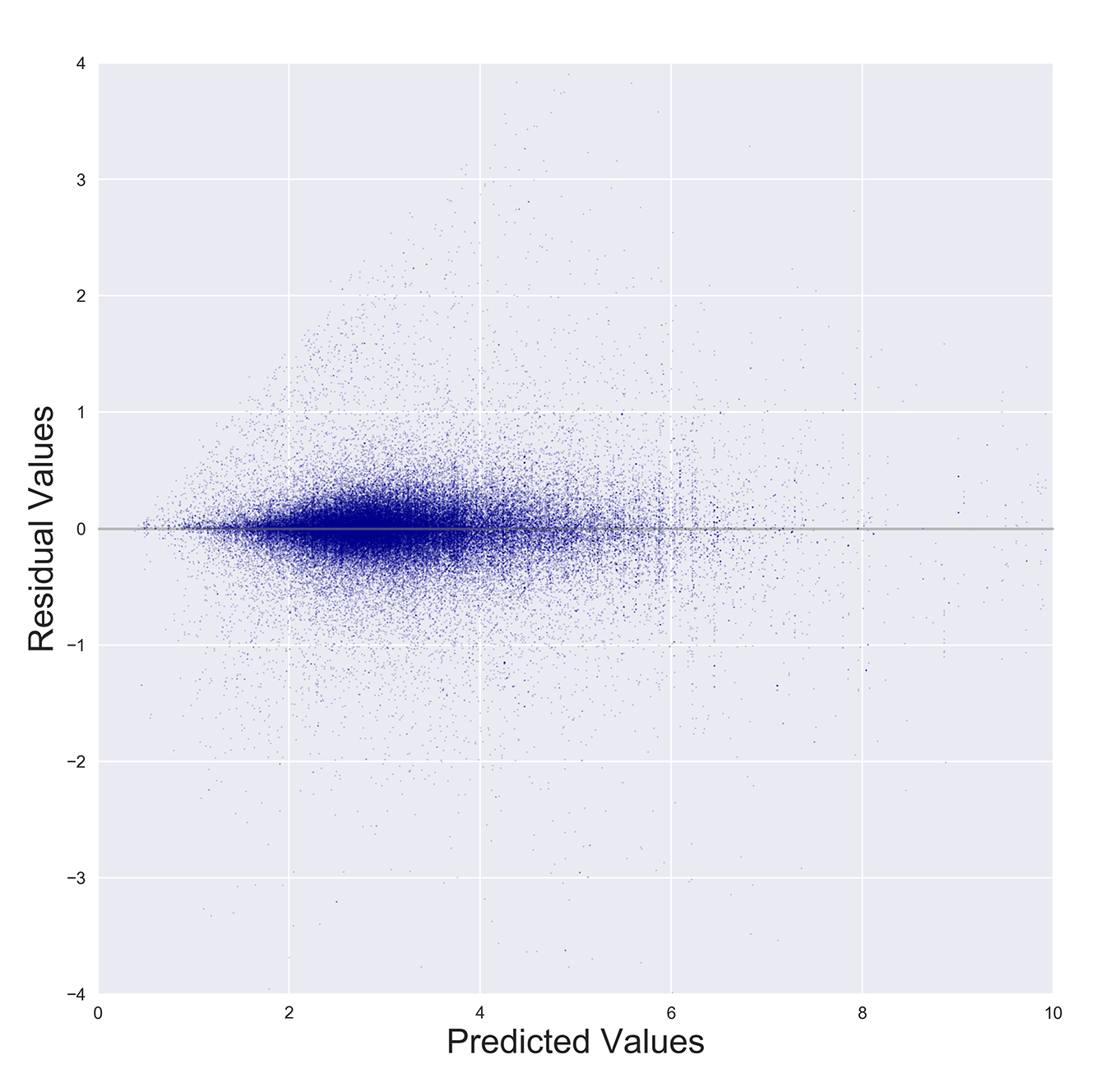}}
\caption{Residuals vs Predicted Values for OLS and Random Forest}
\label{fig:respred}
\end{figure}

Figure \ref{fig:respred} displays the results by plotting the residuals from each model against the predicted values from that model for the testing dataset. This plot can be useful in detecting nonlinear patterns in the errors over the range of the predictions.  In this case the patterns that emerged in Figure \ref{fig:predobs} are still in evidence, with a significant portion of the cloud of points in the OLS plot spreading and drifting downwards at higher predicted values.  By contrast the pattern of residuals vs predicted values is much tighter and does not exhibit a comparable drift in the random forest results.

\begin{figure}[H]
\centering
\subfigure[Ordinary Least Squares]{\includegraphics[width=8cm]{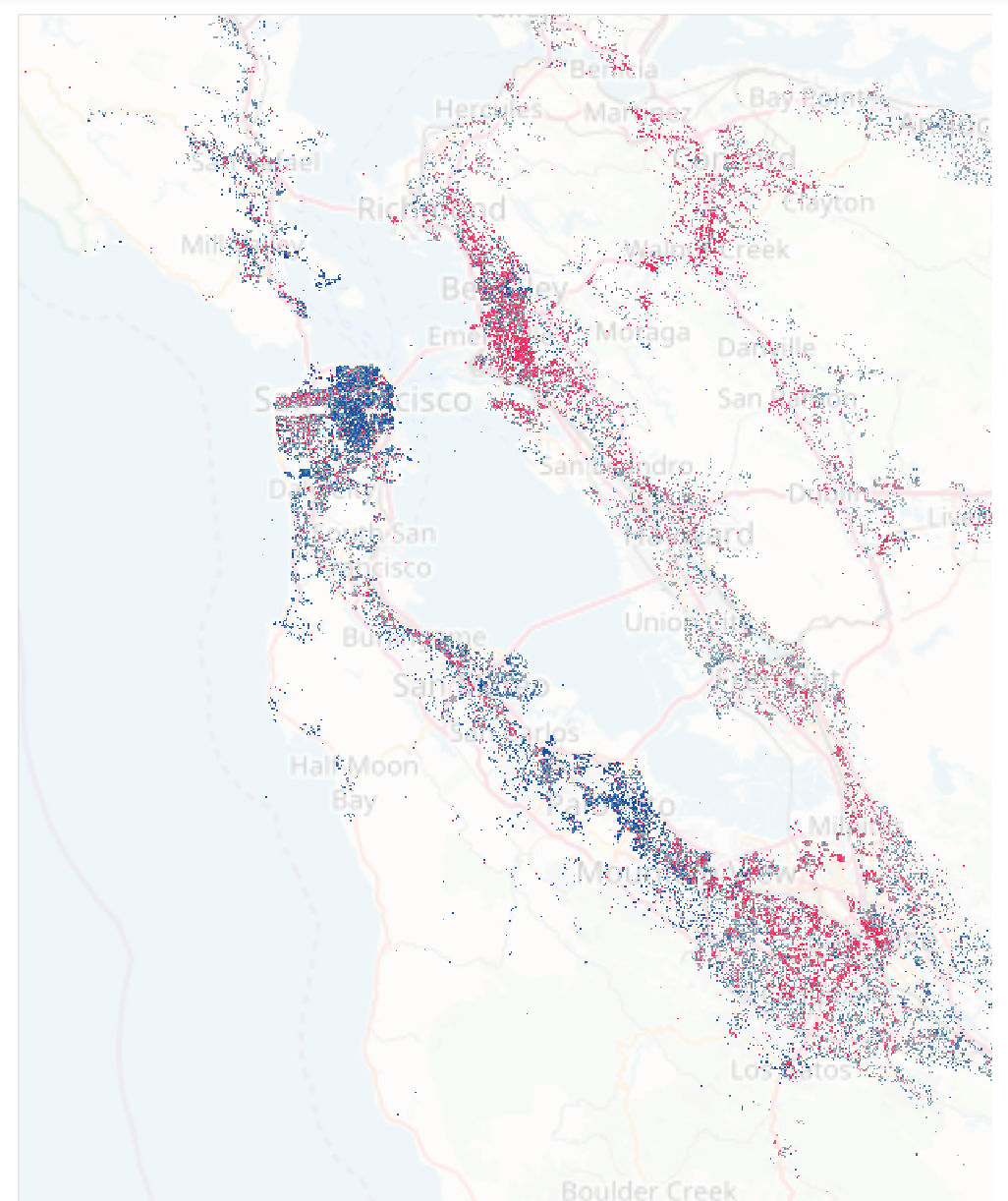}}
\subfigure[Random Forest]{\includegraphics[width=8cm]{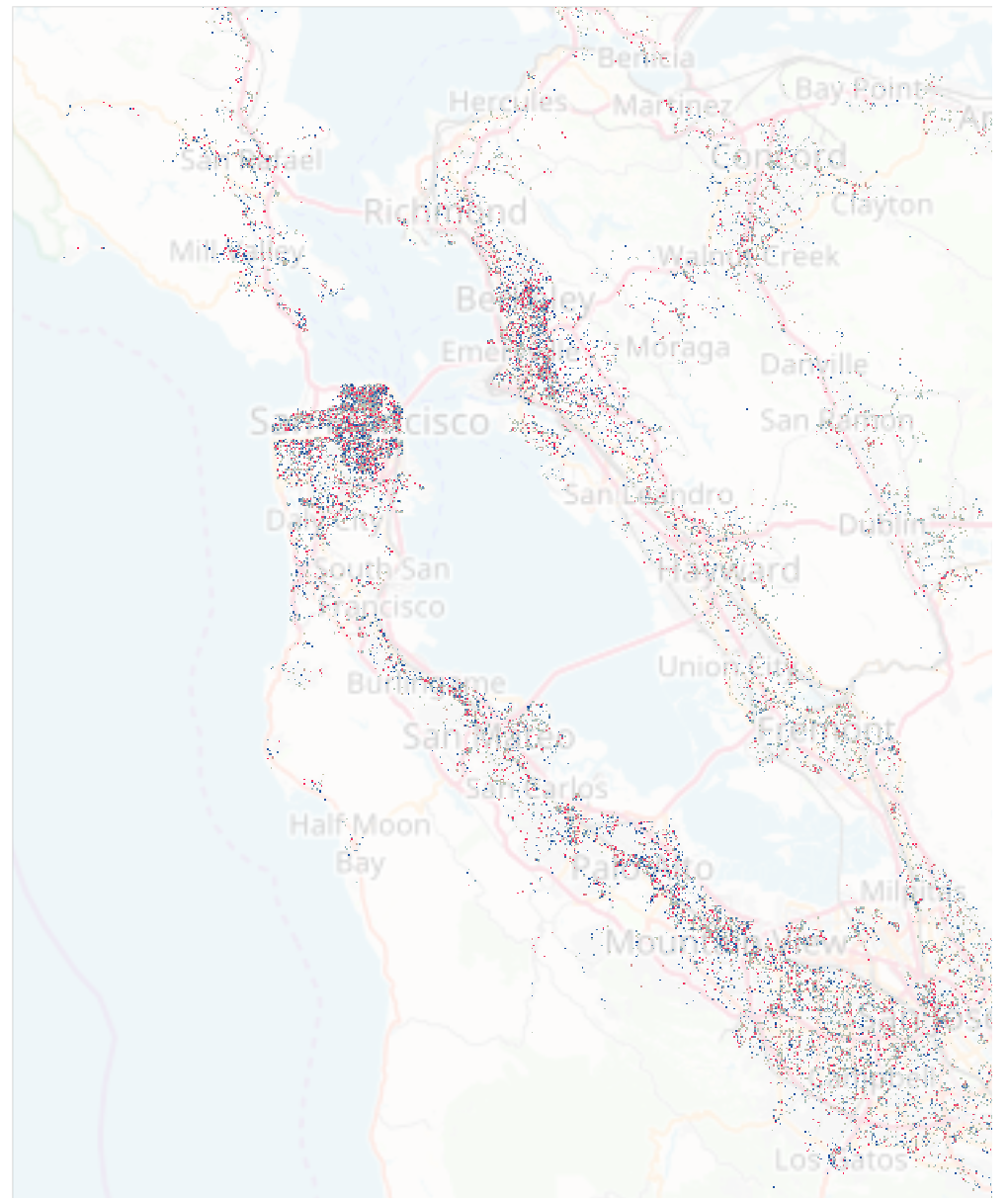}}
\caption{Spatial Pattern of Residuals for OLS and Random Forest}
\label{fig:resmap}
\end{figure}

The final assessment of the residuals from our analysis of the testing dataset is to map the spatial pattern of the residuals from both models to examine them for spatial clustering.  In figure \ref{fig:resmap} we see that there are clusters of underprediction (blues) in the core of San Francisco and in the Silicon Valley area, and over--predictions in parts of Oakland and Berkeley and in parts of the South Bay -- probably reflecting ommitted variables and non-linearities we were not able to capture in the OLS model.  By contrast, the map of residuals from the random forest regression appears random, lacking obvious clustering of under- or over-predictions.

\section{Conclusion}
\label{sec:conclusion}

The most widely used method for undertaking hedonic regression modeling of housing prices and rents, multiple linear regression estimated using ordinary least squares, has a clear advantage over random forest regression in the interpretability of the estimated model coefficients.  If the objective of the application is to evaluate the impact of a specific variable of focus, as is often the case in the hedonic regression literature, then the lower predictive quality of OLS compared to random forest regression is an appropriate trade-off to make.  However, there are other applications in which predictive accuracy is more important than the ability to interpret a specific coefficient, and in such cases, machine learning methods that have higher predictive accuracy than models estimated with OLS are worth considering.  The random forest algorithm was designed in a way that overcomes several limitations of ordinary least squares regression \cite{breiman2001}: 1) it is designed to handle non-linear relationships between the dependent and independent variables; 2) it is invariant to scaling or translation, and 3) it is robust to irrelevant or highly correlated variables. 

Our intended application for the hedonic model is to use the predicted rents as initial values for an integrated land use and transportation model system that is structural.  Hedonic regression models are by construction reduced form models that provide an estimate of the partial influence of a number of independent variables on housing prices or rents, under the assumption that the market is in equilibrium.  However, it may be an appropriate starting point for a structural model in which we predict the demand for each location using discrete choice models with predicted rents on the right hand side, and evaluate the total demand at each location by summing the predicted location probabilities across all choosers at each location, and then iteratively adjusting the rents to account for demand - supply imbalances, to reflect short-term disequilibrium in housing markets.  The second and related use of the rent predictions is as an input into real estate supply models that use pro forma financial models to predict the development profitability or feasibility on a site, given the costs of developing the site, and the expected revenue from constructing a project.  The revenue expectation is heavily informed by predicted rents, and if these predictions are poor, then the quality of the supply model is adversely affected.  In this context, improving the predictive accuracy of the hedonic model is mainly motivated by the need to improve inputs to the demand and supply models, and not so much for its use to evaluate specific coefficients and iterpret them.  This approach puts hedonic regression into a support role within a structural microsimulation of demand and supply. Within contexts such as the use case outlined here, and in most other applications where the accuracy of predictions is more important than the interpretability of a single coefficient, it would appear that machine learning methods such as random forest and the closely related gradient boosting algorithms may have substantial value in improving models.
Finally, we close with a reiteration of our main substantive finding: using only a modest number of local accessibility variables in addition to the square footage of the rental unit as the only unit or building specific attribute, both of the methods used in this study are able to predict rents per square foot with a high degree of accuracy in out of sample predictions.  Local density, social composition, and job accessibility are powerful explanatory factors, and  %
\section{Author Contributions}
\label{sec:contributions}


The authors confirm contribution to the paper as follows: study conception and design: Paul Waddell; 
model development using OLS: Paul Waddell; model development using Machine Learning: 
Arezoo Besharati-Zadeh; analysis and interpretation of results: Paul Waddell and Arezoo
Besharati-Zadeh; draft manuscript preparation: Paul Waddell. All authors reviewed the results and 
approved the final version of the manuscript.
\bibliographystyle{plain}
\bibliography{Waddell-Mendeley}




%
 
\end{document}